\documentclass[aps,prd,preprint,nofootinbib]{revtex4}

\usepackage{graphicx}
\usepackage{psfrag}

\begin{document}
\title{Uncertainties in Estimating the Indirect Production of $B_c$ and Its Excited States
Via Top Quark Decays at CERN LHC}
\author{Xing-Gang Wu\footnote{wuxg@cqu.edu.cn}}
\address{Department of Physics, Chongqing University, Chongqing 400044,
P.R. China}

\begin{abstract}
Main theoretical uncertainties in estimating the indirect production
of $(b\bar{c})$-quarkonium ($B^-_c$ meson and its excited states)
via top quark decays, $t\to (b\bar{c})+c+W^{+}$, are studied within
the non-relativistic QCD framework. It is found that the
dimensionless reduced decay width for a particular
$(b\bar{c})$-quarkonium state,
$\bar\Gamma_{n}=\Gamma_{n}/\Gamma_{t\to W^{+}+b}$, is very sensitive
to the $c$-quark mass, while the uncertainties from the $b$-quark
and $t$-quark masses are small, where $n$ stands for the eight
$(b\bar{c})$-quarkonium states up to ${\cal O}(v^4)$:
$|(b\bar{c})(^1S_0)_1\rangle$, $|(b\bar{c})(^3S_1)_1\rangle$,
$|(b\bar{c})(^1P_1)_1\rangle$, $|(b\bar{c})(^3P_J)_1\rangle$ (with
$J=(1,2,3)$), $|(b\bar{c})(^1S_0)_{8}g\rangle$ and
$|(b\bar{c})(^3S_1)_{8}g\rangle$ respectively. About $10^8$
$t\bar{t}$-pairs shall be produced per year at CERN LHC, if adopting
the assumption that all the higher Fock states decay to the ground
state with $100\%$ probability, then we shall have
$\left(1.038^{+1.353}_{-0.782}\right)\times 10^5$ $B^-_c $ events
per year. So the indirect production provides another important way
to study the properties of $B^-_c$ meson in comparison to that of
the direct hadronic production at CERN LHC. \\


\end{abstract}

\maketitle

\section{Introduction}

The $B_c$ meson is a double heavy quark-antiquark bound state and
carries flavors explicitly, which provides a good platform for a
systematic studies of the $b$ or $c$ quark decays. Since its first
discovery at TEVATRON by CDF collaboration \cite{CDF}, $B_c$ physics
is attracting more and wide interests. Many progresses have been
made for the direct hadronic production of $B_c$ meson at high
energy colliders \cite{chang}, especially, a computer program
BCVEGPY for generating the $B_c$ events has been completed in
Refs.\cite{bcvegpy1,bcvegpy2,bcvegpy3} and has been accepted by
several experimental groups to simulate the $B_c$ events. It has
been estimated with the help of BCVEGPY that about $10^4$ $B_c$
events are expected to be recorded during the first year of the CMS
running with a lepton trigger \cite{cms}, and there are about $10^4$
$B_c$ events with $B_c\to J/\Psi+\pi$ decays in three years of ATLAS
running \cite{atlas}.

On the other hand, the indirect production of $B_c^+$ or $B_c^{-}$,
including its excited states, via $\bar{t}$-decay or $t$-decay may
also provide useful knowledge of these mesons. Without confusing and
for simplifying the statements, later on we will not distinguish
$B^+_c$ and $B^-_c$ (simply call them as $B_c$) and all results for
$B^+_c$ and $B^-_c$ obtained in the paper are symmetric in the
interchange from particle to anti-particle. With a predicted cross
section for top quark pair production hundred times larger than at
TEVATRON and a much higher designed luminosity, e.g. it is expected
that at CERN LHC $\sim 10^8$ $t\,\bar{t}$-pairs can be produced per
year under the luminosity $L=10^{34} cm^{-2}s^{-1}$ \cite{exp}, the
LHC is poised to become a ``top factory". Therefore, the indirect
production of $B_c$ through top quark decays shall provide another
important way to study the properties of $B_c$ meson \cite{qiao}.
Within the non-relativistic QCD (NRQCD) framework \cite{nrqcd}, the
decay channel $t \to (b\bar{c})+c+ W^+$ in leading order (LO)
$\alpha_s$ calculation but with $v^2$-expansion up to $v^4$ has been
recently calculated with the so called `new trace technology'
\cite{ttobc}, where $(b\bar{c})$-quarkonium is in one of the eight
Fock states: the six color-singlet states
$|(b\bar{c})(^1S_0)_1\rangle$, $|(b\bar{c})(^3S_1)_1\rangle$,
$|(b\bar{c})(^1P_1)_1\rangle$ and $|(b\bar{c})(^3P_J)_1\rangle$
(with $J=(1,2,3)$), and two color-octet states
$|(b\bar{c})(^1S_0)_{8} g\rangle$ and
$|(b\bar{c})(^3S_1)_{8}g\rangle$ respectively. It has been argued
that when $10^8$ $t\,\bar{t}$ events per year are produced at LHC,
then it is possible to accumulate about $10^5$ $B_c$ events per year
via $t$-quark decays at LHC. Thus in comparison to that of the
direct hadronic production, there may be some advantages in
$(b\bar{c})$-quarkonium studies via the indirect production due to
the fact that the top quark events shall always be recorded at LHC.

Considering the forthcoming LHC running, and various experimental
feasibility studies of $B_c$ are in progress, it may be interesting
to know the theoretical uncertainties quantitatively in estimating
of $B_c$ production. The uncertainties of the direct hadronic
production of $B_c$ through its dominant gluon-gluon fusion
mechanism has been studied in Refs.\cite{diruncern,octet}, while the
present paper is served to study the uncertainties of the indirect
mechanism through the decay channel $t \to (b\bar{c})+c+ W^+$. These
two cases are compensate to each other and may be useful for
experimental studies. At the present, we shall restrict ourselves to
examine the uncertainties at the lowest order, because the
next-to-leading order (NLO) calculation cannot be available soon due
to its complicatedness. For definiteness, we shall examine the main
uncertainties that are caused by the value of the $t$-quark mass,
the values of the bound state parameters $m_c$ and $m_b$, and the
choice of the renormalization scale $Q^2$.

The remainder of the paper is organized as follows. Section II gives
the calculation technology for the indirect production of
$(b\bar{c})$-quarkonium states through the top quark decays. Section
III is devoted to present the numerical results and to discuss the
corresponding uncertainties with the help of the formulae given in
Ref.\cite{ttobc} . And section IV is reserved for a summary.

\section{calculation technology}

Within the non-relativistic QCD (NRQCD) frame work \cite{nrqcd}, the
dimensionless reduced decay width for the production of
$(b\bar{c})$-quarkonium through the channel $t(p_0)\to
(b\bar{c})(p_1)+c(p_2)+ W^+(p_3)$ takes the following factorization
form:
\begin{equation}
\bar\Gamma =\sum_{n}\bar\Gamma_n=\sum_{n}\left[\frac{1}{\Gamma_{t\to
W^{+}+b}} H_n(t\to(b\bar{c})+c+ W^+)\times\frac{\langle{\cal
O}_n\rangle} {N_{col}}\right] ,
\end{equation}
where $\bar\Gamma_n$ stands for the reduced decay width for a
particular $(b\bar{c})$-quarkonium state, and the sum is over all
the $(b\bar{c})$-quarkonium states up to ${\cal O}(v^4)$, which
includes six color singlets $|(b\bar{c})(^1S_0)_1\rangle$,
$|(b\bar{c})(^3S_1)_1\rangle$, $|(b\bar{c})(^1P_1)_1\rangle$ and
$|(b\bar{c})(^3P_J)_1\rangle$ (with $J=(1,2,3)$), and two color
octets $|(b\bar{c})(^1S_0)_{8}g\rangle$ and
$|(b\bar{c})(^3S_1)_{8}g\rangle$ respectively. $N_{col}$ refers to
the number of colors, $n$ stands for the involved states of
$(b\bar{c})$-quarkonium. $N_{col}=1$ for singlets and
$N_{col}=N_c^2-1$ for octets. $\langle{\cal O}_n\rangle$ stands for
the decay matrix element that can be related with the wave function
at zero $R_S(0)$ or the derivative of the radial wave function at
origin $R'_P(0)$ through the saturation approximation \cite{nrqcd}.
The overall factor $1/\Gamma_{t\to W^{+}+b}$ is introduced to cut
off the uncertainty from the electroweak coupling. The decay width
of the two body decay process $t(p_1)\to b(p_2)+ W^+(p_3)$ that is
dominant for the $t$-quark decays can be written as
\begin{equation}
\Gamma_{t\to W^{+}+b}=\frac{G_F m_t^2|\mathbf{\vec{p}_2}|}
{4\sqrt{2\pi}}\left[(1-y^2)^2+x^2(1+y^2-2x^2)\right] ,
\end{equation}
where $|\mathbf{\vec{p}_2}|=\frac{m_t}{2}
\sqrt{(1-(x-y)^2)(1-(x+y)^2)}$, $x=m_{w}/m_t$ and $y=m_{b}/m_t$.

\begin{figure}
\centering
\includegraphics[width=0.60\textwidth]{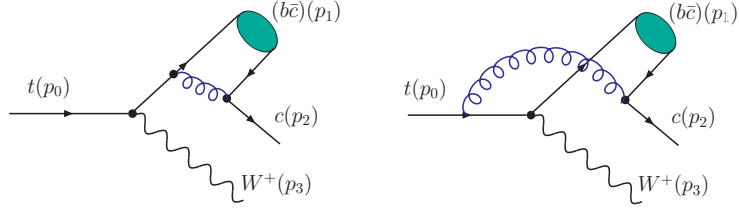}
\caption{Feynman diagrams for the indirect production of
$(b\bar{c})$-quarkonium through top quark decays. } \label{feyn}
\end{figure}

As shown in Fig.(\ref{feyn}), there are two Feynman diagrams for the
concerned process $t(p_0)\to (b\bar{c})(p_1)+c(p_2)+ W^+(p_3)$. Due
to the involved massive quarks, the calculation of the process is
very complicated and lengthy, to simplify the calculation, we have
improved a so called `new trace technology' to calculate the process
\cite{ttobc}. Under such approach, we first arrange the whole
amplitude into several orthogonal sub-amplitudes $M_{ss'}$ according
to the spins of the $t$-quark ($s'$) and $c$-quark ($s$), and then
do the trace of the Dirac $\gamma$ matrix strings at the amplitude
level by properly dealing with the massive spinors, which results in
explicit series over some independent Lorentz-structures, and
finally, we obtain the square of the amplitude. All the necessary
formulae together with its subtle points for the square of the hard
scattering amplitude $H_n(t\to(b\bar{c})+c+ W^+)$ can be found in
Ref.\cite{ttobc}, so we shall only present the main results here and
the interesting reader may turn to Ref.\cite{ttobc} for more
detailed calculation technology.

The involved color-singlet and color-octet matrix elements provide
systematical errors for the NRQCD framework itself. Their values can
be determined by global fitting of the experimental data or directly
related to the wave functions at the zero point $R_S(0)$ (or the
derivative of the wave function at the zero point $R'_P(0)$) derived
from certain potential models for the color-singlet case, some
potential models can be found in Ref.\cite{pot1,pot2,pot3,pot4}. A
model dependent analysis of $R_S(0)$ and $R'_P(0)$ can be found in
Ref.\cite{quigg}, where the spectrum of $B_c$ under the Cornell
potential \cite{pot1}, the Buchm${\rm \ddot{u}}$ller-Tye potential
\cite{pot2}, the power-law potential \cite{pot3} and the logarithmic
potential \cite{pot4} have been discussed respectively in their
discussions, which shows that $|R_S(0)|^2\in [1.508,1.710] {\rm
GeV}^3$ and $|R'_P(0)|^2\in [0.201, 0.327] {\rm GeV}^5$
\footnote{Since the Cornell potential has stronger singularity in
spatially smaller states \cite{quigg}, so we do not include its
corresponding values for $R_S(0)$ and $R'_P(0)$.}. Since the
model-dependent $R_S(0)$ and $R'_P(0)$ emerge as overall factors and
their uncertainties can be conveniently discussed when we know their
possible ranges well, so we shall not discuss such uncertainties in
the present paper. More explicitly, we shall fix their values to be:
$|R_S(0)|^2=1.642\; {\rm GeV}^3$ and $|R'_P(0)|^2=0.201\; {\rm
GeV}^5$, which is derived under the Buchm${\rm \ddot{u}}$ller-Tye
potential \cite{quigg}. Secondly, although we do not know the exact
values of the two decay color-octet matrix elements, $\langle
b\bar{c}(^1S_0)_8|{\cal O}_8(^1S_0)| b\bar{c}(^1S_0)_8 \rangle$ and
$\langle b\bar{c}(^3S_1)_8|{\cal O}_8(^3S_1)| b\bar{c}(^3S_1)_8
\rangle$, we know that they are one order in $v^2$ higher than the
$S$-wave color-singlet matrix elements according to NRQCD scale
rule. More specifically, based on the velocity scale rule
\cite{nrqcd}, we have
\begin{equation}
\langle b\bar{c}(^1S_0)_8|{\cal O}_8(^1S_0)| b\bar{c}(^1S_0)_8
\rangle \simeq \Delta_S(v)^2\cdot \langle b\bar{c}(^1S_0)_1|{\cal
O}_1(^1S_0)| b\bar{c}(^1S_0)_1 \rangle
\end{equation}
and
\begin{equation}
\langle b\bar{c}(^3S_1)_8|{\cal O}_8(^3S_1)| b\bar{c}(^3S_1)_8
\rangle \simeq \Delta_S(v)^2\cdot \langle b\bar{c}(^3S_1)_1|{\cal
O}_1(^3S_1)| b\bar{c}(^3S_1)_1 \rangle\,,
\end{equation}
where the second equation comes from the vacuum-saturation
approximation. $\Delta_S(v)$ is of order $v^2$ or so, and we take it
to be within the region of 0.10 to 0.30, which is in consistent with
the identification: $\Delta_S(v)\sim\alpha_s(Mv)$ and has covered
the possible variation due to the different ways to obtain the wave
functions at the origin ($S$-wave) and the first derivative of the
wave functions at the origin ($P$-wave) etc.

In addition to the color-singlet and color-octet matrix elements,
the quark mass values $m_t$, $m_c$ and $m_b$ also `generate'
uncertainties for the hadronic production. At present, these
parameters cannot be completely fixed by fitting the available data
of the heavy quarkonium. Furthermore, since the
$(b\bar{c})$-quarkonium state is the non-relativistic and
weak-binding bound state, we approximately have $M_{B_c}=m_b + m_c$,
which also is the requirement from the gauge invariance of the hard
scattering amplitude.

To choose the renormalization scale $Q^2$ is a tricky problem for
the estimates of the LO pQCD calculation. If $Q^2$ is chosen
properly, the results may be quite accurate. In the present case
with three-body final state, there is ambiguity in choosing the
renormalization scale $Q^2$ and various choices of $Q^2$ would
generate quite different results. Such kind of ambiguity cannot be
justified by the LO calculation itself, so we take it as the
uncertainty of the LO calculation, although when the NLO calculation
of the subprocess is available, the uncertainty will become under
control a lot. While the NLO calculation is very complicated and it
cannot be available in the foreseeable future, so here we take $Q^2$
as the possible characteristic momentum of the hard subprocess being
squared. According to the factorization formulae, the running of
$\alpha_s$ should be of leading logarithm order, and the energy
scale $Q^2$ appearing in the calculation should be taken as one of
the possible characteristic energy scales of the hard subprocess. As
a default choice, we take $Q^2=4m_c^2$, since the intermediate gluon
should be hard enough to produce a $c$ and $\bar{c}$ pair as shown
in Fig.(\ref{feyn}).

\section{Numerical results and discussions}

\begin{table}
\begin{center}
\caption{Reduced decay width $\bar\Gamma_n$ for the indirect
production of $B_c$ through top quark decays with varying $m_c$,
$m_b$ and $m_t$, where $n$ stands for a particular $(cb)$-quarkonium
state.} \vspace{0.5cm}
\begin{tabular}{|c||c||c|c||c|c||c|c||}
\hline ~~~$m_c$ (GeV)~~~& ~~~$1.5$~~~&  ~~~$1.2$~~~ &
~~~$1.8$~~~ & \multicolumn{2}{|c||}{$1.5$} & \multicolumn{2}{|c||}{$1.5$}  \\
\hline $m_b$ (GeV)& $4.9$ & \multicolumn{2}{|c||}{$4.9$}& ~~~$4.5$~~~ & ~~~$5.3$~~~
& \multicolumn{2}{|c||}{$4.9$}  \\
\hline $m_t$ (GeV)& ~~~$172.5$~~~&\multicolumn{2}{|c||}{$172.5$} &
\multicolumn{2}{|c||}{$172.5$} & ~~~$170.$~~~& ~~~$175.$~~~ \\
\hline\hline $\bar\Gamma_{(^1S_0)_1}(\times 10^4)$ & 3.590 &7.095 & 2.053& 3.605 & 3.573& 3.580& 3.600 \\
\hline $\bar\Gamma_{(^3S_1)_1}(\times 10^4)$ & 4.975 & 10.530& 2.690& 4.877&5.065 &4.959 & 4.991 \\
\hline\hline $\bar\Gamma_{(^1P_1)_1}(\times 10^5)$ &3.543 & 10.704& 1.449& 3.628& 3.470& 3.526&3.559 \\
\hline $\bar\Gamma_{(^3P_0)_1}(\times 10^5)$ &2.110&5.294 & 1.009 &2.301 &1.952 &2.106 &2.114  \\
\hline $\bar\Gamma_{(^3P_1)_1}(\times 10^5)$ &4.390&12.528 & 1.883 &4.585 & 4.224&4.372 & 4.407\\
\hline $\bar\Gamma_{(^3P_2)_1}(\times 10^5)$ &4.718&15.642 & 1.765&4.652 & 4.774& 4.693&4.743 \\
\hline
\end{tabular}
\label{tabm}
\end{center}
\end{table}

Firstly, we study the uncertainties of $m_t$, $m_c$ and $m_b$ in `a
factorizable way' by fixing the renormalization scale $Q^2=4m_c^2$.
For instance, when focussing on the uncertainties from $m_c$, we let
it be a basic `input' parameter varying in a possible range
$m_c=1.5\pm0.3\; {\rm GeV}$ with all the other factors, including
the $t$-quark mass, $b$-quark mass and {\it etc.} being fixed to
their center values. The Particle Data Group value for the top quark
mass is $m_t=172.5\pm 2.7\; {\rm GeV}$ \cite{pdg}. And the $b$-quark
mass $m_b$ varies within the region of $m_b=4.9\pm0.4\; {\rm GeV}$.
The reduced decay width $\bar\Gamma_n$ for the indirect production
of $B_c$ through top quark decays with varying $m_c$, $m_b$ and
$m_t$ is shown in TAB.\ref{tabm}, where $n$ stands for a particular
color singlet $(cb)$-quarkonium state. The results for the two
$S$-wave color octet  can be conveniently obtained from that of
color singlet $S$-wave $(cb)$-quarkonium states and by setting
$\Delta_S(v)\in[0.10, 0.30]$. The second column of TAB.\ref{tabm} is
for the center values of all these parameters, the third and fourth
columns setting the upper and the lower limit for $m_c$ varying
within the region of $[1.2, 1.8]$ GeV, the fifth and the sixth
columns setting the upper and the lower limit for $m_b$ varying
within the region of $[4.5, 5.3]$ GeV, and the seventh and eighth
columns setting the upper and the lower limit for $m_t$ varying
within the region of $[170., 175.]$ GeV respectively.

From TAB.\ref{tabm}, it is found that the reduced decay width
$\bar\Gamma_n$ is very sensitive to $m_c$. $\bar\Gamma_n$ decreases
with the increment of $m_c$, and more definitely, when $m_c$
increase by steps of $0.1{\rm GeV}$, $\bar\Gamma_n$ decreases by
$10\% \sim 20\%$ for $S$-wave states and by $25\%\sim 35 \%$ for
$P$-wave states. This condition is similar to the direct hadronic
production \cite{diruncern}, which is caused by the fact that a
larger $m_c$ leads to a smaller allowed phase space. Summing up all
the mentioned Fock states' contribution, we obtain $\sum_n
\bar\Gamma_n=\left(1.038^{+1.324}_{-0.498}\right)\times10^{-3}$ for
$m_c\in[1.2,1.8] {\rm GeV}$ and $\Delta_S(v)\in[0.10, 0.30]$, where
the center value is for $m_c=1.5 {\rm GeV}$, $m_b=4.9 {\rm GeV}$,
$m_t=172.5 {\rm GeV}$ and $\Delta_S(v)=0.2$.

The reduced decay width $\bar\Gamma_n$ slightly decreases with the
increment of $m_b$ for $n=|(b\bar{c})(^1S_0)_1\rangle$,
$n=|(b\bar{c})(^1S_0)_{8}g\rangle$, $|(b\bar{c})(^1P_1)_1\rangle$,
$|(^3P_0)_1\rangle$ and $|(b\bar{c})(^3P_1)_1\rangle$ respectively,
but increases with the increment of $m_b$ for
$n=|(b\bar{c})(^3S_1)_{1}\rangle$, $|(b\bar{c})(^3S_1)_{8}g\rangle$
and $|(b\bar{c})(^3P_2)_1\rangle$ respectively. As for the direct
hadronic production of $B_c$, since there is a $b$-quark jet in the
final state, so the production shall always decrease with the
increment of $b$-quark mass \cite{diruncern}. While for the present
case, there is no such $b$-quark jet in the final state, so the
condition is slightly different. Further more, it is found that when
$m_b$ increase by steps of $0.2{\rm GeV}$, the uncertainties is less
than $1\%$. Summing up all the mentioned Fock states' contribution,
we obtain $\sum_n \bar\Gamma_n=\left(1.038^{+0.037}_{-0.022}\right)
\times 10^{-3}$ for $m_b\in[4.5,5.3] {\rm GeV}$ and
$\Delta_S(v)\in[0.10, 0.30]$.

The reduced decay width $\bar\Gamma_n$ slightly increases with the
increment of $m_t$, which is due to the larger phase space for a
larger $m_t$. To check the results of Ref.\cite{ttobc}, we also
calculate the results for $m_t=176 {\rm GeV}$, which shows a good
agreement with those of Ref.\cite{ttobc}. Summing up all the
mentioned Fock states' contribution, we obtain $\sum_n
\bar\Gamma_n=\left(1.038^{+0.046}_{-0.029}\right) \times10^{-3}$ for
$m_t\in[170,175] {\rm GeV}$ and $\Delta_S(v)\in[0.1,0.3]$.

\begin{table}
\begin{center}
\caption{Reduced decay width $\bar\Gamma_n$ for the indirect
production of $B_c$ through top quark decays with three typical
renormalization scale $Q^2$, where $n$ stands for a particular
color-singlet $(cb)$-quarkonium state.}
\begin{tabular}{|c||c|c|c|}
\hline ~~$m_c$ ({\rm GeV})~~ &  ~~~ $Q^2=E_{B_c}^2$ ~~~
&~~~$Q^2=4m_b^2$~~~ & ~~~$Q^2=4m_c^2$~~~ \\
\hline\hline $\bar\Gamma_{(^1S_0)_1}(\times 10^4)$  & 2.125 &
1.738 & 3.590   \\
\hline $\bar\Gamma_{(^3S_1)_1}(\times 10^4)$ & 2.875 & 2.408
& 4.975  \\
\hline\hline $\bar\Gamma_{(^1P_1)_1}(\times 10^5)$ & 2.040
& 1.715 & 3.543   \\
\hline $\bar\Gamma_{(^3P_0)_1}(\times 10^5)$ &  1.177 & 1.022
& 2.110  \\
\hline $\bar\Gamma_{(^3P_1)_1}(\times 10^5)$  & 2.546  &
2.125 & 4.390 \\
\hline $\bar\Gamma_{(^3P_2)_1}(\times 10^5)$  &  2.700  & 2.284
 & 4.718 \\
\hline
\end{tabular}
\label{tabQ2}
\end{center}
\end{table}

Secondly, we study the uncertainties caused by the various choices
of $Q^2$, where for consistency, the leading order $\alpha_s$
running is adopted, i.e. $\alpha_s(Q^2)=4\pi/[(11-\frac{2}{3}n_f)
\ln (Q^{2}/\Lambda^2_{QCD})]$, where $n_f=3$ and $\Lambda_{QCD}=200
{\rm MeV}$. We choose three typical renormalization scale $Q^2$:
Type A: $Q^2=4m^2_c$; Type B: $Q^2=4m_b^2$; Type C: $Q^2=E_{B_c}^2$,
where $E_{B_c}$ stands for the $B_c$ meson energy in the top quark
rest frame, and by setting $s_2=(p_2+p_3)^2$ for the channel
$t(p_0)\to (b\bar{c})(p_1)+c(p_2)+ W^+(p_3)$, we have
$E_{B_c}=(m_t^2+M_{B_c}^2-s_2)/(2m_t)$. The uncertainties for the
reduced decay width $\bar\Gamma_n$ with three typical choices of
$Q^2$ are given in TAB.\ref{tabQ2}, where $m_c=1.5 GeV$, $m_b=4.9
GeV$ and $m_t=172.5 GeV$. It is found that the reduced width for
$Q^2=4m_b^2$ is only about half of that of $Q^2=4m_c^2$, which is a
comparatively large effect.

\begin{figure}
\centering
\includegraphics[width=0.45\textwidth]{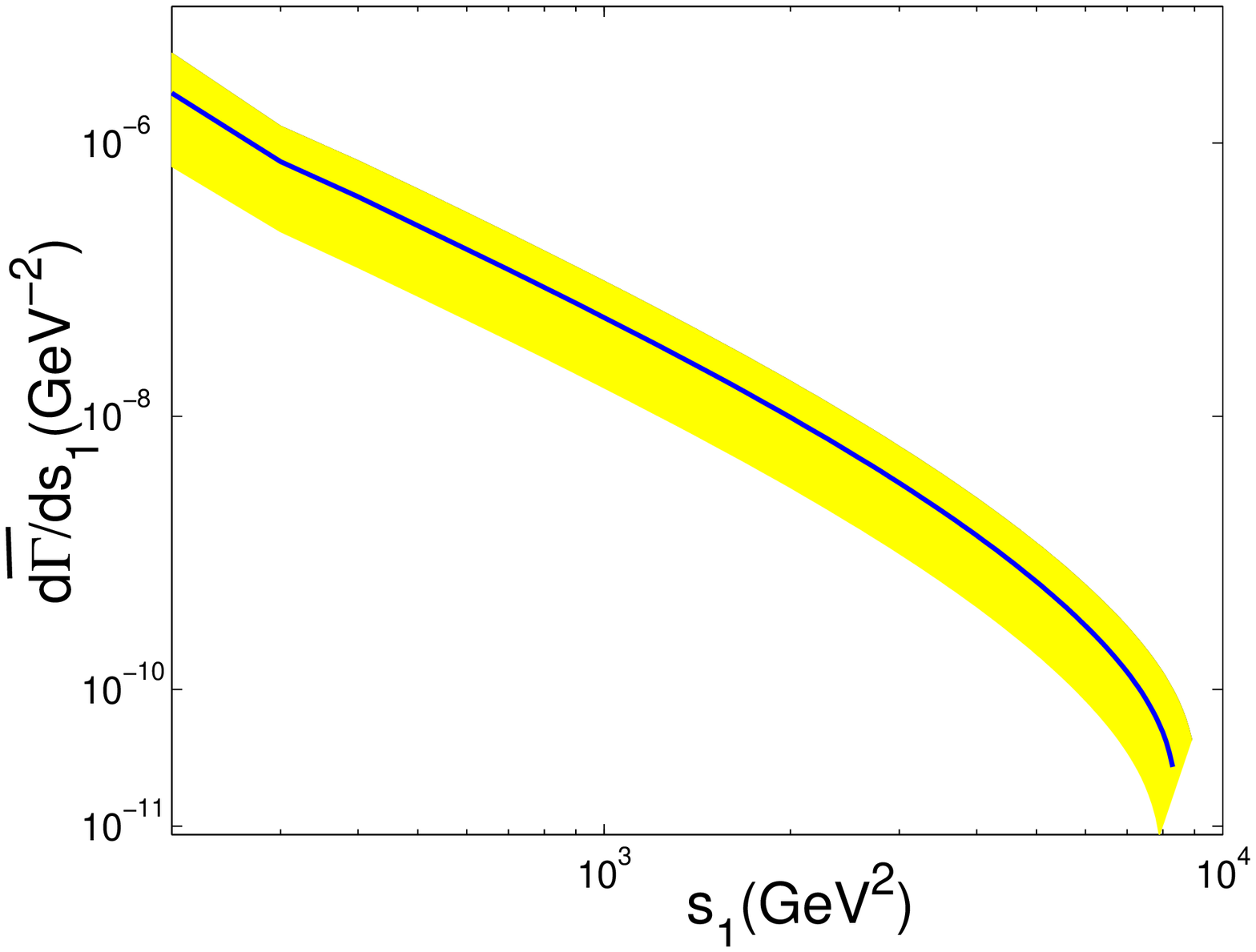}
\includegraphics[width=0.45\textwidth]{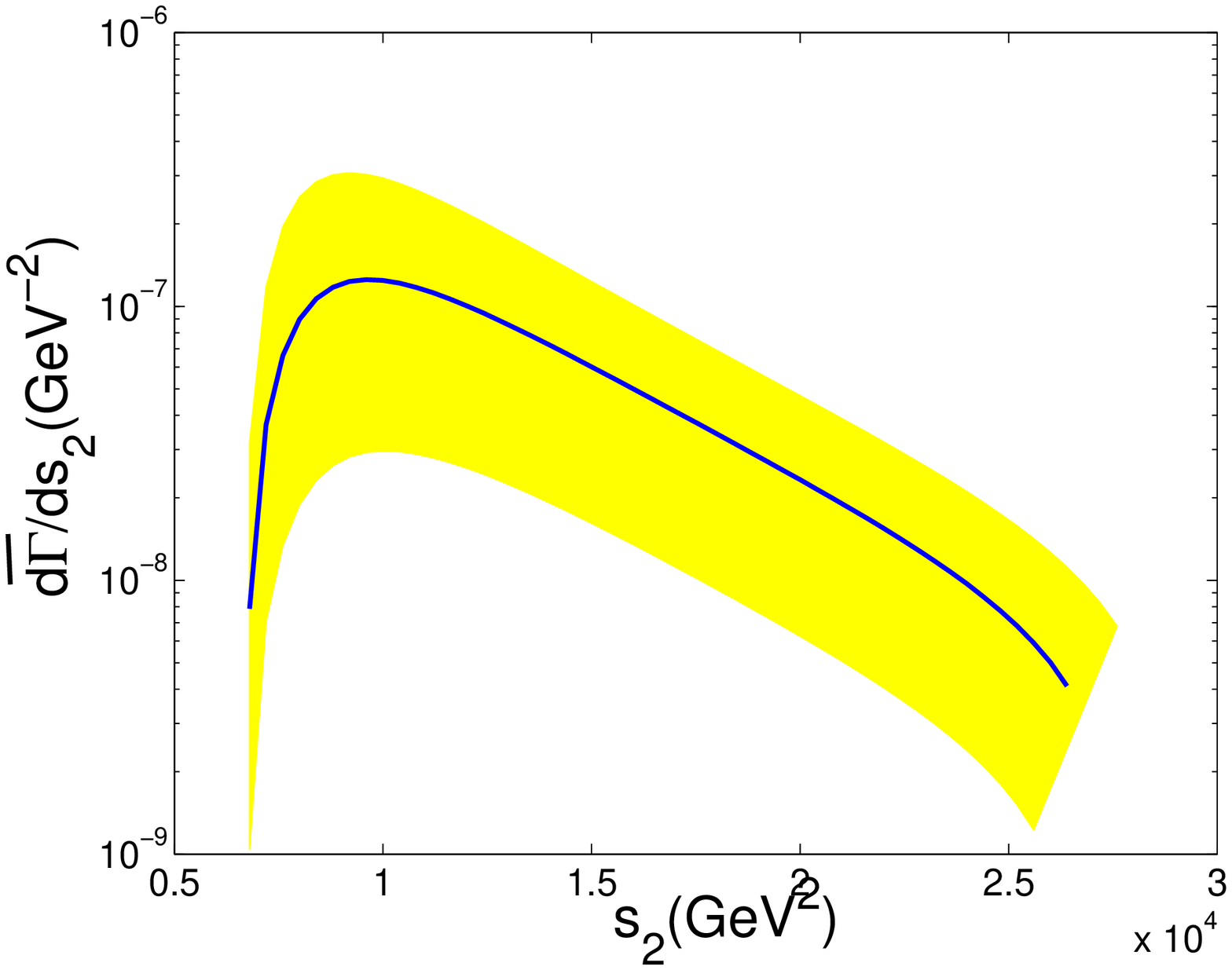}
\caption{Differential distributions $d\bar\Gamma/ds_1$ and
$d\bar\Gamma/ds_2$. The shaded shows the uncertainty and the solid
line is for the center value with $m_c=1.5 {\rm GeV}$, $m_b=4.9 {\rm
GeV}$ and $m_t=172.5 {\rm GeV}$.} \label{diss}
\end{figure}

\begin{figure}
\centering
\includegraphics[width=0.45\textwidth]{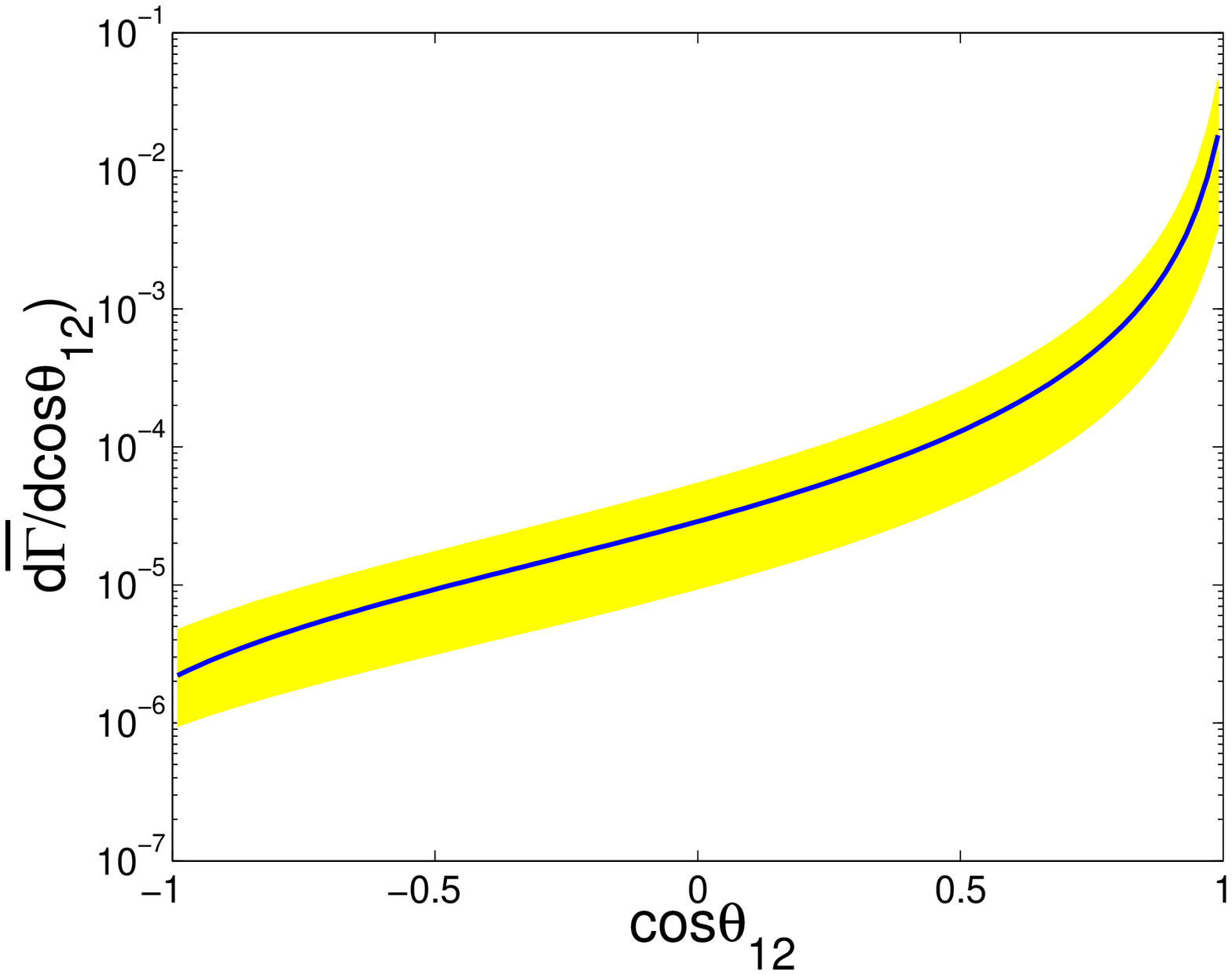}
\includegraphics[width=0.45\textwidth]{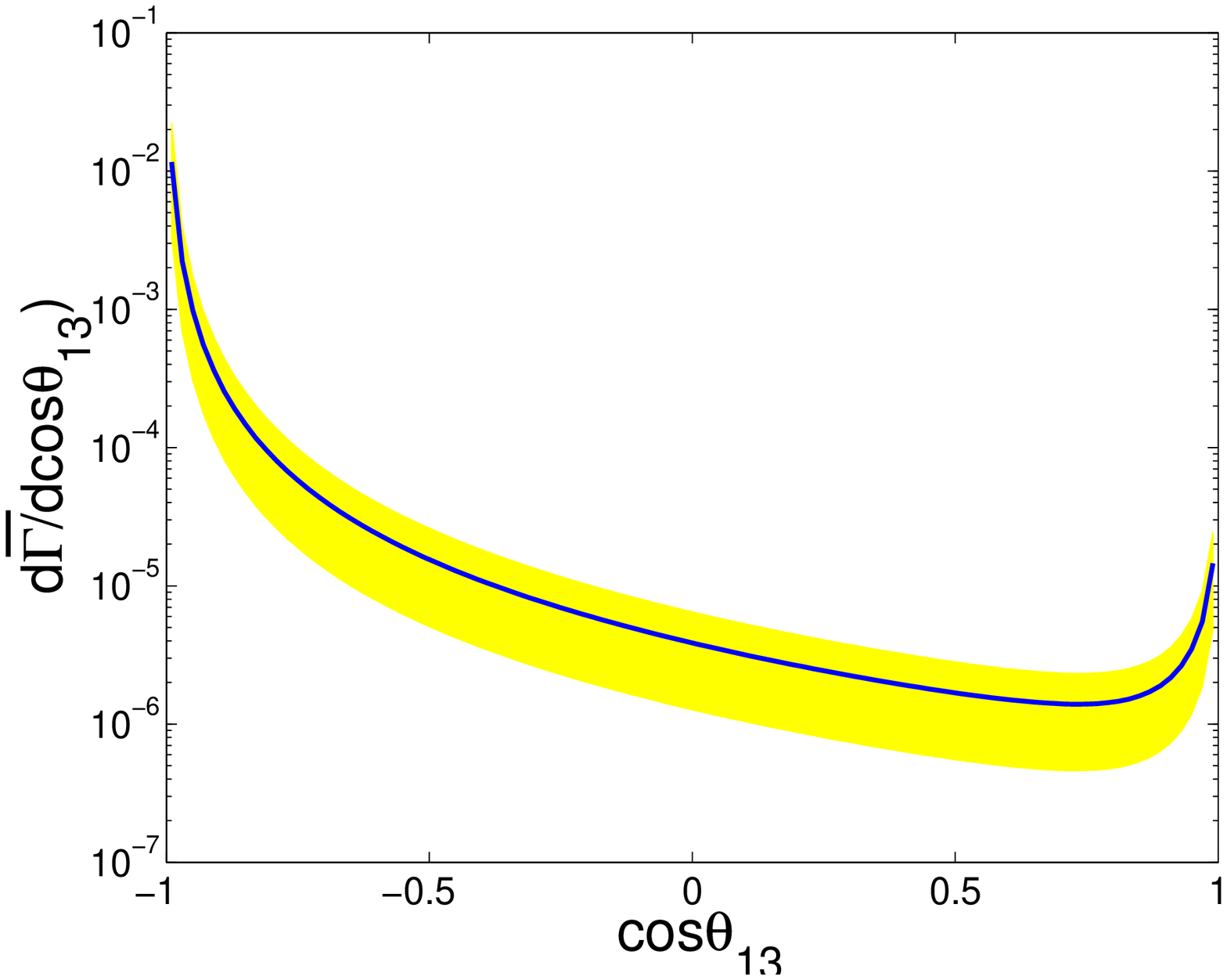}
\caption{Differential distributions $d\bar\Gamma/dcos\theta_{12}$
and $d\bar\Gamma/dcos\theta_{13}$. The shaded shows the uncertainty
and the solid line is for the center value with $m_c=1.5 {\rm GeV}$,
$m_b=4.9 {\rm GeV}$ and $m_t=172.5 {\rm GeV}$.} \label{discos}
\end{figure}

Finally, we discuss the combined effects of all the above mentioned
uncertainty sources by varying $m_c\in[1.2,1.8] {\rm GeV}$,
$m_b\in[4.5,5.3] {\rm GeV}$, $m_t\in[170,175] {\rm GeV}$ and by
taking one of the three typical renormalization scales
simultaneously, where all the mentioned Fock states' contributions
shall be summed up. Additionally, the value of $m_b+m_c$ can not be
too small, as has been found both experimentally and theoretically
that the mass of the ground state $(b\bar{c})$-quarkonium is around
$6.30{\rm GeV}$ \cite{cdfmass,allison}, so we imply $m_b+m_c\gtrsim
6.20{\rm GeV}$ as an extra constraints. Summing up all the mentioned
Fock states' contribution, we obtain $\sum_n
\bar\Gamma_n=\left(1.038^{+1.353}_{-0.782}\right) \times10^{-3}$.
Let us show some more characteristics of the decay
$t\to(b\bar{c})+c+ W^+$. The differential distributions of the
reduced decay width versus the invariant masses $s_1=(p_1 +p_2)^2$
and $s_2=(p_2 +p_3)^2$, i.e. $d\bar\Gamma/ds_1$ and
$d\bar\Gamma/ds_2$ are shown in Fig.(\ref{diss}). While the
differential distributions of the reduced decay width versus
$\cos\theta_{13}$ and $\cos\theta_{12}$, i.e.
$d\bar\Gamma/d\cos\theta_{12}$ and $d\bar\Gamma/d\cos\theta_{13}$
are shown in Fig.(\ref{discos}), where $\theta_{13}$ is the angle
between $\vec{p}_1$ and $\vec{p}_3$, and $\theta_{12}$ is the angle
between $\vec{p}_1$ and $\vec{p}_2$ respectively in the $t$-quark
rest frame ($\vec{p}_0=0$). In drawing the curves, all the mentioned
Fock states' contribution have been summed up for convenience. The
shaded band shows the corresponding uncertainty, with the upper edge
of the band is obtained by setting $m_c=1.2 {\rm GeV}$, $m_b=5.0
{\rm GeV}$, $m_t=175 {\rm GeV}$ and $Q^2=4m_c^2$ and the lower edge
of the band is obtained by setting $m_c=1.8 {\rm GeV}$, $m_b=5.3
{\rm GeV}$, $m_t=170 {\rm GeV}$ and $Q^2=4m_b^2$ , and the center
solid line is for $m_c=1.5 {\rm GeV}$, $m_b=4.9 {\rm GeV}$,
$m_t=172.5 {\rm GeV}$ and $Q^2=4m_c^2$.

\section{Summary}

In the paper we have presented quantitative studies on the main
uncertainties in estimating the indirect production of the
$(b\bar{c})$-quarkonium via top quark decays, $t\to
(b\bar{c})+c+W^{+}$. It is found that the reduced decay width
$\bar\Gamma_n$ is very sensitive to the $c$-quark mass, while the
uncertainty from the $b$-quark and $t$-quark masses are small. The
renormalization scale also affects the decay width to a certain
degree. A comparative study on the similarity and difference of the
direct and indirect production has also been presented in due
places. About $10^8$ $t\bar{t}$-pairs shall be produced per year at
CERN LHC, if adopting the assumption that all the higher Fock states
decay to the ground state with $100\%$ probability, then we may have
$\left(1.038^{+1.353}_{-0.782}\right)\times 10^5$ $B^-_c (B^+_c)$
events per year. So the indirect production is another important way
to study the properties of $B_c$ meson in comparison to that of the
direct hadronic production at LHC. Further more, the contribution
from the $P$-wave states together with the two color-octet Fock
states can be about $20\%$ in total, so the $P$-wave production
itself is worthwhile to study the possibility of directly measuring
the $P$-wave $B_c$ states.

\begin{center}
\section*{Acknowledgements}
\end{center}

This work was supported in part by Natural Science Foundation
Project of CQ CSTC under grant number 2008BB0298 and Natural Science
Foundation of China under grant number 10805082, the National Basic
Research Programme of China under Grant No 2003CB716300, and by the
grant from the Chinese Academy of Engineering Physics under the
grant numbers: 2008T0401 and 2008T0402. \\


\begin{thebibliography}{s2}

\bibitem{CDF} CDF Collaboraten, F. Abe, {\em et al.}, Phys. Rev. Lett.
{\bf 81}, 2432 (1998); Phys. Rev. {\bf D58}, 112004 (1998).

\bibitem{chang} Chao-Hsi Chang, Int.J.Mod.Phys. A{\bf 21},
777(2006) and references therein.

\bibitem{bcvegpy1} Chao-Hsi Chang, Chafik Driouich, Paula Eerola and Xing-Gang Wu,
Comput. Phys. Commun. {\bf 159}, 192(2004).

\bibitem{bcvegpy2} Chao-Hsi Chang, Jian-Xiong Wang and Xing-Gang Wu, Comput. Phys.
Commun. {\bf 174}, 241(2006).

\bibitem{bcvegpy3} Chao-Hsi Chang, Jian-Xiong Wang and Xing-Gang Wu, Comput. Phys.
Commun. {\bf 175}, 624(2006).

\bibitem{cms} S.H. Zhang, A.A. Belkov, S. Shulga and G.M. Chen,
chin.Phys.Lett.{\bf 21}, 2380(2004).

\bibitem{atlas} V. Kartvelishvili, {\it for the ATLAS
Collaboration}, Nucl.Phys. B (Proc.Suppl) {\bf 164}, 161(2007).

\bibitem{exp} N. Kidonakis and R. Vogt, Int. J. Mod. Phys. A{\bf 20}, 3171, 2005;
F. Hubaut, {\it et al.}, ATLAS collaboration, hep-ex/0605029; V.
Barger and R.J. Phillips, Report No. MAD/PH/789, 1993.

\bibitem{qiao} C.F. Qiao, C.S. Li and K.T. Chao, Phys.Rev. D{\bf
54}, 5606(1996).

\bibitem{nrqcd} G.T. Bodwin, E. Braaten and G.P. Lepage,
Phys. Rev. D {\bf 51}, 1125 (1995); Erratum Phys. Rev. D {\bf 55},
5853 (1997).

\bibitem{ttobc} Chao-Hsi Chang, Jian-Xiong Wang and Xing-Gang Wu,
Phys.Rev. D{\bf 77}, 014022(2008).

\bibitem{diruncern} Chao-Hsi Chang and Xing-Gang Wu, Eur.Phys.J. C{\bf 38}, 267(2004).

\bibitem{octet} Chao-Hsi Chang, Cong-Feng Qiao, Jian-Xiong Wang and
Xing-Gang Wu, Phys.Rev. D{\bf 71}, 074012(2005).

\bibitem{pot1}  E. Eichten, K. Gottfried, T. Kinoshita, K.D. Lane and
T.M. Yan, Phys.Rev. D{\bf 17}, 3090(1978); ibid. {\bf 21},
313(E)(1980); ibid.{\bf 21}, 203(1980).

\bibitem{pot2} W. Buchm${\rm \ddot{u}}$ller and S.-H.H. Tye, Phys.Rev. D{\bf 24},
132(1981).

\bibitem{pot3} A. Martin, Phys.Lett. B{\bf 93}, 338(1980).

\bibitem{pot4} C. Quigg and J.L. Rosner, Phys.Lett. B{\bf 71},
153(1977).

\bibitem{quigg} E.J. Eichten and C. Quigg, Phys.Rev. D{\bf 49},
5845(1994).

\bibitem{pdg} W.M. Yao {\it etal.}, J.Phys. G {\bf 33}, 1(2006).

\bibitem{cdfmass} CDF Collaboraten, A. Abulencia, {\em et al.}, Phys. Rev. Lett.
{\bf 96}, 082002(2006).

\bibitem{allison} I.F. Allison {\it et al.}, Phys.Rev. Lett.{\bf
94}, 172001(2005), and reference therein.

\end{thebibliography}
\end{document}